\documentclass[twocolumn,showpacs,preprintnumbers,amsmath,amssymb,prb,superscriptaddress]{revtex4-1}

\usepackage{amssymb}
\usepackage{graphicx}
\usepackage{bm}
\usepackage{natbib}

\begin{document}
\title{Symmetry and correlations underlying Hidden Order in URu$_2$Si$_2$}

\author{Nicholas P. Butch}
\email{nicholas.butch@nist.gov}
\affiliation{Center for Nanophysics and Advanced Materials, Department of Physics, University of Maryland, College Park, MD 20742}
\affiliation{NIST Center for Neutron Research, National Institute of Standards and Technology, 100 Bureau Drive, Gaithersburg, MD 20899}
\affiliation{Lawrence Livermore National Laboratory, 7000 East Ave., Livermore, CA 94550}
\author{Michael E. Manley}
\affiliation{Oak Ridge National Laboratory, Oak Ridge, Tennessee 37831}
\author{Jason R. Jeffries}
\affiliation{Lawrence Livermore National Laboratory, 7000 East Ave., Livermore, CA 94550}
\author{Marc Janoschek}
\affiliation{Los Alamos National Laboratory, Los Alamos, New Mexico 87545}
\affiliation{Department of Physics, University of California - San Diego, 9500 Gilman Drive, La Jolla, CA 92093}
\author{Kevin Huang}
\author{M. Brian Maple}
\affiliation{Department of Physics, University of California - San Diego, 9500 Gilman Drive, La Jolla, CA 92093}
\author{Ayman H. Said}
\author{Bogdan M. Leu}
\affiliation{Advanced Photon Source, Argonne National Laboratory, 9700 S. Cass Ave., Argonne, IL 60439}
\author{Jeffrey W. Lynn}
\affiliation{NIST Center for Neutron Research, National Institute of Standards and Technology, 100 Bureau Drive, Gaithersburg, MD 20899}
\date{\today}

\begin{abstract}
We experimentally investigate the symmetry in the Hidden Order (HO) phase of intermetallic URu$_2$Si$_2$ by mapping the lattice and magnetic excitations via inelastic neutron and x-ray scattering measurements in the HO and high-temperature paramagnetic phases. At all temperatures the excitations respect the zone edges of the body-centered tetragonal paramagnetic phase, showing no signs of reduced spatial symmetry, even in the HO phase. The magnetic excitations originate from transitions between hybridized bands and track the Fermi surface, whose feature are corroborated by the phonon measurements. Due to a large hybridization energy scale, a full uranium moment persists in the HO phase, consistent with a lack of observed crystal-field-split states. Our results are inconsistent with local order parameter models and the behavior of typical density waves. We suggest that an order parameter that does not break spatial symmetry would naturally explain these characteristics.

\end{abstract}

\maketitle

The underlying cause of a large entropy change at the Hidden Order (HO) transition temperature $T_\mathrm{HO}=17.5$~K in URu$_2$Si$_2$ remains a mystery, despite many experimental and theoretical developments over more than a quarter century since its discovery \cite{Schlabitz86,Palstra85,Maple86}. The HO phase develops out of a complicated correlated electronic paramagnetic state built of interacting itinerant and localized uranium \emph{f}-electron states. Moreover, it is unstable to an unconventional superconducting ground state. Identifying the HO parameter is among the most persistent and thought-provoking challenges facing condensed matter physics.

Experiments cannot conclusively identify a symmetry-breaking order parameter to account for the configurational entropy change measured at the ordering temperature. The shape of the specific heat anomaly at $T_\mathrm{HO}$ resembles the second-order Bardeen-Cooper-Schriefer superconducting transition, suggesting that an energy gap is created in the itinerant electron states \cite{Palstra85,Maple86}. Energy gaps in the electronic states are also inferred from many other measurements \cite{Levallois11,Schmidt10,Aynajian10,Park12,Liu11,Shirer13}, but recent findings suggest that these features really develop at temperatures greater than $T_\mathrm{HO}$, so it is likely that these gaps are associated with the development of local-itinerant electron correlations starting at much higher temperatures. These correlations are further associated with magnetic excitations that develop at high temperature but become gapped and dispersing in the HO phase \cite{Broholm91,Wiebe07}. As to the nature of the order parameter, early neutron diffraction identified A-type antiferromagnetic (AFM) order below $T_\mathrm{HO}$, but the small measured moment is incompatible with the large entropy release \cite{Broholm87}. This sample-dependent moment actually arises due to defects \cite{Amitsuka02,Takagi07} that stabilize puddles of an inhomogeneous large-moment antiferromagnetic (AFM) phase \cite{Amitsuka99,Matsuda01}, which evolves into a bulk phase above a first-order phase transition at 0.8~GPa \cite{Butch10}. Although an extrinsic origin is not yet universally accepted, any AFM-type moment intrinsically associated with the HO phase must be very small, aligned out of plane \cite{Das13,Metoki13,Ross14}, and can be removed by light chemical substitution \cite{Williams12}. Similarly, x-ray diffraction measurements indicate that there is no change in crystal symmetry through $T_\mathrm{HO}$ \cite{Kernavanois99}. There is also no evidence for local rotational symmetry breaking on Ru and Si sites \cite{Mito13} or antiferroquadrupolar order \cite{Amitsuka10}. Nonetheless, experiment \cite{Okazaki11,Meng13} and theory \cite{Harima10,Chandra13,Oppeneer10,Fujimoto11,Ikeda12} continue to suggest that the HO transition involves a reduction of lattice symmetry.

To address this issue, we looked for signs of incipient symmetry-breaking by measuring the magnetic and lattice excitations of URu$_2$Si$_2$ in the HO and paramagnetic phases to energies as high as 30~meV across much of reciprocal space. We draw several concrete conclusions from this extensive study. Temperature-dependent magnetic scattering that follows the development of electronic correlations is concentrated along zone edges. Overall, the magnetic and lattice excitations always respect the symmetry of the high-temperature paramagnetic phase, contrary to prevailing ideas that the HO phase shares the broken lattice symmetry of the pressure-induced AFM phase. Our data are thus inconsistent with theories invoking a primary uranium-based antiferromultipolar order parameter. Moreover, we show that the shape of the magnetic dispersion in reciprocal space, and the temperature-dependence of certain phonon modes, provide evidence for the Fermi surfaces involved in hybridization between itinerant electrons and localized \emph{f}-states. We also find that the full uranium $J$ is responsible for the magnetic excitations in the HO phase. Our results neatly demonstrate the dual itinerant/local electron nature of the correlated electron state from which the HO phase emerges and constrain possible HO models.

\subsection{Experiment}

Neutron scattering measurements were performed on a 7~g single crystal of URu$_2$Si$_2$ that was synthesized via the Czochralski technique in a continuously-gettered, tetra-arc furnace and subsequently annealed. The sample exhibits a small out-of-plane ordered moment of 0.016(1)~$\mu_B$/U and negligible in-plane moment smaller than $2 \times 10^{-3} \mu_B$/U \cite{Das13}. Inelastic neutron scattering measurements were carried out on the BT-7 thermal triple axis spectrometer at the NIST Center for Neutron Research \cite{Lynn12}. Temperature was controlled by a closed-cycle refrigerator. Inelastic scans were measured at constant wavevector $Q$ and varying energy transfer $E$. Typical scattering conditions were 50' - 25' - 50' - 120' collimation with 14.7~meV final energy. Energy resolution was approximately 1.2~meV full-width-half-max at the elastic position. Data were collected in both $a-a$ (basal) plane and $a-c$ plane geometries. Polarized neutron scattering measurements were performed using a $^3$He-based apparatus \cite{Chen11} with open - 50' - 80' - 120' collimation, a vertical guide field, and 14.7~meV final energy. A flipping ratio of 60 was determined from the nuclear $(2,0,0)$ reflection.

Time-of-flight inelastic neutron scattering measurements were performed on the NG-4 disc-chopper spectrometer at the NIST Center for Neutron Research \cite{Copley03}. Temperature was controlled by a closed-cycle refrigerator. The instrument was run in low-resolution mode with incident energy 13.09~meV and scattering in the $a-a$ plane. Data were collected over 180$^\circ$ of sample rotation. Energy resolution ranged from 0.77~meV
at elastic scattering to 0.4~meV at 10~meV transfer. Data analysis was performed using the DAVE software suite \cite{Azuah09}.

Inelastic x-ray scattering measurements on a single crystal with lateral dimensions $0.3$~mm and thickness $0.015$~mm were carried out on the HERIX spectrometer \cite{Toellner11,Said11} at Sector~30 of the Advanced Photon Source using 23.7~keV incident energy photons, with an energy resolution of 1.5~meV. The sample was taken from a large single crystal that exhibits a small out-of-plane moment of 0.011~$\mu_B$/U \cite{Butch10}. Approximate x-ray spot size on sample was $35\times15$~$\mu$m$^2$. A pressure of 2.0~GPa was applied via a diamond anvil cell using a 4:1~methanol/ethanol pressure medium, while temperature was controlled using a closed-cycle refrigerator. Ruby fluorescence was used for manometry. Measurements were performed along the $(h,h,0)$ direction.

Throughout this paper, error bars associated with measurements and fits correspond to one standard deviation unless otherwise noted. Error bars not plotted are smaller than the plotted points.

\subsection{Magnetic excitations}

\subsubsection{The Hidden Order phase}

\begin{figure}
\begin{center}
\includegraphics[width=3.4in]{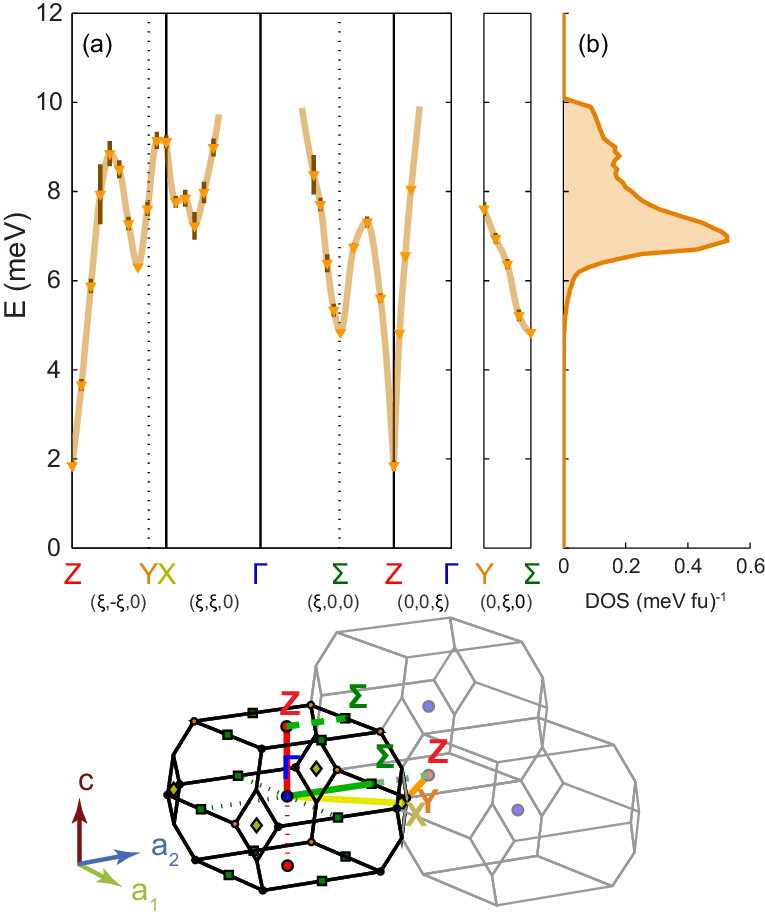}
\end{center}
\caption{Magnetic excitations in the HO phase. a) Dispersion of magnetic excitations at 2.6~K and b) corresponding magnetic density of states (DOS). The narrow bandwidth of the excitations reflects their origin in a hybridized \emph{f}-band. Data were taken on BT-7. High-symmetry paths in the Brillouin zone of the BCT unit cell correspond to the wavevectors plotted above.}
\label{magDOS}
\end{figure}

For orientation, a reciprocal space map of the body-centered tetragonal (BCT) lattice of URu$_2$Si$_2$ shows the Brillouin zone (BZ) in Fig.~\ref{magDOS}, identifying the high-symmetry points and lines along which the dispersions have been plotted. Points of particular interest are BZ center $\pmb{\Gamma}$, horizontal face center \textbf{X}, vertical face center \textbf{Z} and horizontal edge center $\pmb{\Sigma}$, as well as corners \textbf{Y} of horizontal zone faces. Historically, these reciprocal-lattice points have been labeled using a simple tetragonal (ST) coordinate system, such that \textbf{Z} = \{1,0,0\} and $\pmb{\Sigma}$ = \{$q_1$,0,0\} where $q_1 = \frac{1}{2}(1 + \frac{a^2}{c^2}) \approx 0.6$. Note that the horizontal path $\pmb{\Gamma}$-$\pmb{\Sigma}$ extends to \textbf{Z}, which sits on the vertical zone boundary between adjacent BZs that are offset along the $c$-axis. When discussing directions, we remain consistent with the literature and refer to the $a = (1,0,0)$ and $c = (0,0,1)$ axes of the ST unit cell with lattice parameters $\mathrm{a} = 4.13$~{\AA} and $\mathrm{c} = 9.58$~{\AA}.

The magnetic excitations along important reciprocal space directions, as measured on BT-7, are summarized in Fig.~\ref{magDOS}a. Consistent with previous reports, the global dispersion minimum with a value of 1.8~meV is found at \textbf{Z}, while a broad local minimum with a value of 4.8~meV is centered at $\pmb{\Sigma}$ \cite{Broholm87,Bourdarot10,Wiebe07}. The commensurate \textbf{Z} point corresponds to the ordering vector in the AFM phase and features the smallest energy gap. However, unlike typical low-$E$ transverse magnons, the magnetic excitations near \textbf{Z} are longitudinal, with a spin orientation along the out-of-plane magnetic easy axis of the system. Meanwhile, the magnetic excitations at $\pmb{\Sigma}$ feature a larger energy gap, the opening of which can quantitatively account for the entropy change at $T_\mathrm{HO}$ \cite{Wiebe07}, constituting an important signature of the HO transition. We do not observe magnetic excitations above 10~meV, nor near $\pmb{\Gamma}$. The appreciable coverage of reciprocal space by our measurements allows a \emph{model-independent} numerical interpolation of the dispersion inside the entire BZ, from which a magnetic density of states (DOS) is calculated, shown in Fig.~\ref{magDOS}b. This DOS peaks strongly near 7~meV, demonstrating that a narrow 6-8~meV range of $E$ dominates the excitation spectrum in the HO phase. This energy scale corresponds to the coherence temperature of $80$~K \cite{Palstra85,Maple86,Schlabitz86}.

\begin{figure}
\begin{center}
\includegraphics[width=3.4in]{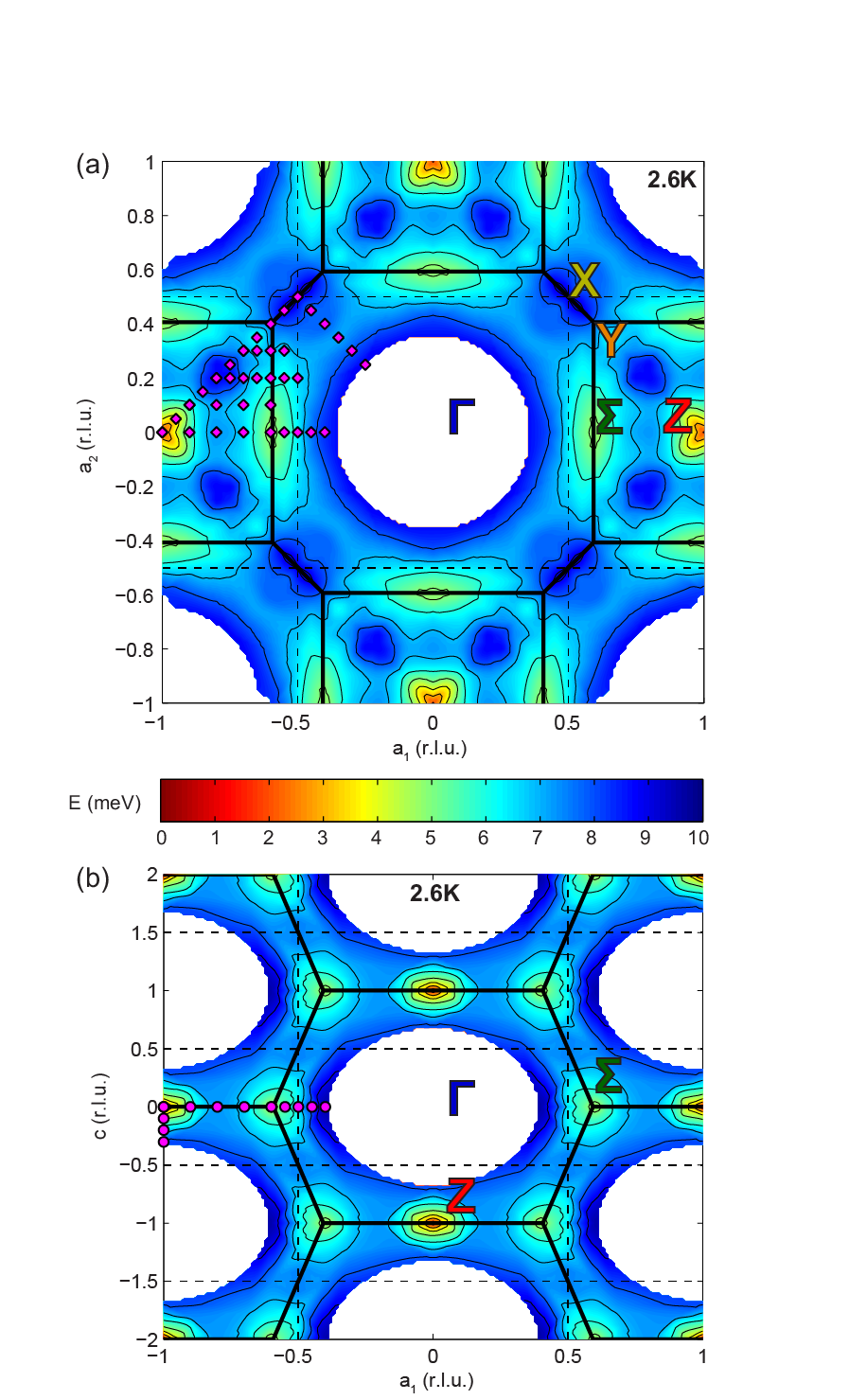}
\end{center}
\caption{Magnetic dispersion in the HO phase in a) the basal plane and b) the $a$-$c$ plane. The energies are identified in the color bar. Minima exist at the \textbf{Z} and $\pmb{\Sigma}$ points and the magnetic excitations are experimentally unobserved in proximity to the $\pmb{\Gamma}$ point. In a) it is apparent that the excitations near the \textbf{Z} and especially the $\pmb{\Sigma}$ minima disperse anisotropically. In b) the  dispersion is more circular or ellipsoidal about the minima. Along both projections, the magnetic excitations trace the boundaries of the BCT reciprocal lattice (in heavy black). For reference, dotted lines represent the ST lattice of the AFM unit cell. Magenta points denote the $q$ at which the excitations were measured. Data were tesselated to create the 2D plots and $c$-axis dispersion data near $\pmb{\Sigma}$ from Ref.~\cite{Wiebe07} were used to fill in figure b. Note that these are \emph{not} intensity plots. Data were taken on BT-7.}
\label{2Dmagdisps}
\end{figure}

\begin{figure}
\begin{center}
\includegraphics[width=3.4in]{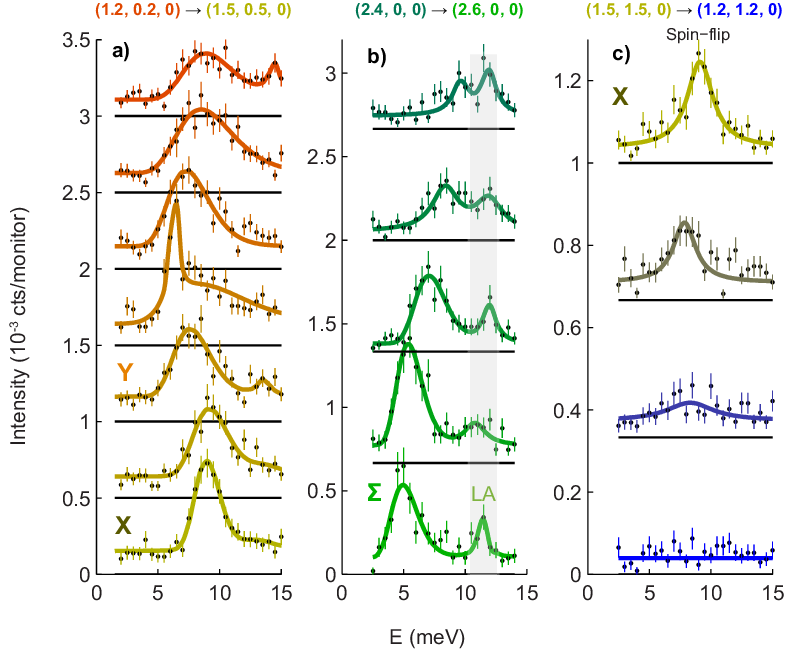}
\end{center}
\caption{Magnetic excitations in the HO phase along selected cuts in the basal plane. They cover portions of a) \textbf{Z}-\textbf{X}, b) $\pmb{\Gamma}$-$\pmb{\Sigma}$, and c) \textbf{X}-$\pmb{\Gamma}$. Data sets are offset vertically for clarity. Initial and final $q$-values are listed at the top of each figure pane, and data progress down from initial to final $q$ in even spacings. These plots give a sense of how the magnetic excitation disperses and weakens away from the zone edges. Note that the data in c) are from polarized neutron scattering, which is only sensitive to magnetic scattering in the spin-flip channel. The gray band in b) delineates scattering from the flat part of a longitudinal acoustic phonon mode. Data were taken on BT-7. Magnetic excitations were fit to log-normal functions as described in the text, while phonons were fit to Lorentzians.}
\label{HOmag}
\end{figure}

Figure~\ref{2Dmagdisps}a is a reciprocal space map composed from the energies of maximum inelastic magnetic intensity determined from numerous constant-$Q$ (examples plotted in Fig.~\ref{HOmag}) scans and plotted in a symmetrized reduced zone scheme. Looking away from the commonly-studied reduced wavevectors $q$, the magnetic dispersion in the HO phase is strikingly anisotropic in the tetragonal basal plane. The local minimum at $\pmb{\Sigma}$ is sharp along the $\pmb{\Gamma}$-\textbf{Z} direction, which led Wiebe and coworkers to model the spectrum as a gapped cone of high velocity incommensurate spin excitations \cite{Wiebe07}. This description is appropriate for the dispersion in the $a-c$ plane, as shown in Fig.~\ref{2Dmagdisps}b. Note, however, that the dispersion in the perpendicular $\pmb{\Sigma}$-\textbf{Y} direction is much weaker, and hence the excitation velocity is much lower (Fig.~\ref{2Dmagdisps}a). Yet, this anisotropy arises naturally from BCT symmetry, because $\pmb{\Sigma}$ sits on the zone boundary (heavy black lines in Fig.~\ref{2Dmagdisps}) separating one BZ from the shared top/bottom face of its immediate neighbors. The perpendicular dispersion actually represents a zone edge mode extending to the square corner \textbf{Y}. To quantify the dispersion asymmetry, we approximate it as purely quadratic near $\pmb{\Sigma}$ for simplicity: the corresponding coefficients are $520 \pm 30$~meV-{\AA}$^{2}$ towards $\pmb{\Gamma}$ and $93 \pm 2$~meV-{\AA}$^{2}$ towards \textbf{Y} (Fig.~\ref{magDOS}), a difference of roughly a factor of 5. Although there is no symmetry relating the $\pmb{\Sigma}$-$\pmb{\Gamma}$ and $\pmb{\Sigma}$-\textbf{Z} directions, the excitations are symmetric within a 0.1 reciprocal lattice unit (r.l.u.) window $\approx 0.15$~{\AA}$^{-1}$ of $\pmb{\Sigma}$. The magnetic excitations thus sit on a line that follows the $\pmb{\Sigma}$-\textbf{Y} zone edge. Near the corner \textbf{Y}, this line curves slightly inward toward \textbf{Z}, and the local minimum sits 0.15~r.l.u. off of \textbf{Y}. Note that the resultant ring is not circular, in that the minima are not equidistant from \textbf{Z}, in contrast to earlier suggestions \cite{Buyers94}. In all directions, as the excitations disperse upward in $E$, they broaden and weaken, except near the \textbf{X} point.

The magnetic excitations in the vicinity of \textbf{X} have received little attention since early work by Broholm et al. \cite{Broholm91}. The path \textbf{Y}-\textbf{X} traverses a zone edge, tracing a line across a face shared by adjacent BZs. Following this path towards \textbf{Z}, the peaks due to magnetic scattering become asymmetric and broaden greatly (Fig.~\ref{HOmag}a), similar in appearance to those observed near $\pmb{\Sigma}$ (Fig.~\ref{HOmag}b) and \textbf{Z} \cite{Broholm91,Wiebe07,Bourdarot10}. The linewidths far exceed the instrument resolution, estimated to have a half-width of 0.75~meV at these $E$ transfers, which points to an intrinsic origin. As we discuss below, this asymmetry naturally arises in the context of interband scattering. The lineshapes are for simplicity fit to log-normal functions, but a sharp peak at the dispersion minimum demonstrates the limitations of this phenomenological treatment. We tentatively treat this sharp peak as intrinsic because its area is necessary to keep the integrated intensity consistent with neighboring $Q$-points. In the perpendicular direction \textbf{X}-$\pmb{\Gamma}$, the magnetic excitations disperse initially downward and then upward in $E$, hitting a local minimum approximately $\frac{1}{3}$ of the way towards $\pmb{\Gamma}$ (Fig.~\ref{HOmag}c). The magnetic nature of these excitations, which are similar in $E$ to the phonons at \textbf{X} has been confirmed via polarized neutron scattering (Fig.~\ref{HOmag}c). The magnetic excitations are impossible to track past a point halfway toward $\pmb{\Gamma}$. A similar difficulty is encountered along the $\pmb{\Gamma}$-$\pmb{\Sigma}$ and $\pmb{\Gamma}$-\textbf{Z} branches, such that the dispersions near $\pmb{\Gamma}$ in any direction remain experimentally undefined.

The excitations in the $a-c$ plane, shown in Fig.~\ref{2Dmagdisps}b, are consistent with earlier inelastic neutron data \cite{Wiebe07}. We first focus on the $\pmb{\Sigma}$-centered dispersion, which is roughly isotropic in the $a-c$ plane. The $\pmb{\Sigma}$-centered excitations are stacked in a vertical zig-zag that coincides with an overlay of the BCT reciprocal lattice (black lines), underscoring the obvious correspondence between the lattice edges and the magnetic excitations. Many previous studies have shown that the magnetic excitations at \textbf{Z} are intense and long-lived \cite{Broholm91,Wiebe07,Bourdarot05}. However, these excitations weaken and are very difficult to track beyond only about 0.15~r.l.u. away, which is true along both $a-$ and $c-$ directions. This differs from the extended $q$ range of the $\pmb{\Sigma}$-\textbf{Y}-\textbf{X} excitations, and the distinct nature of the \textbf{Z} excitations becomes clear when comparing their intensities. Overall, it is clear that the HO magnetic dispersion consists entirely of zone-boundary modes that broaden and weaken away from the zone edges and faces. Most of the intensity is in the square \textbf{Z}-centered faces, which are connected through branches of weaker excitations both in- and out-of-plane, forming a 3D network. We emphasize that these excitations do not respect ST symmetry, which has a BZ shaped like a right square prism.

\begin{figure}
\begin{center}
\includegraphics[width=3.4in]{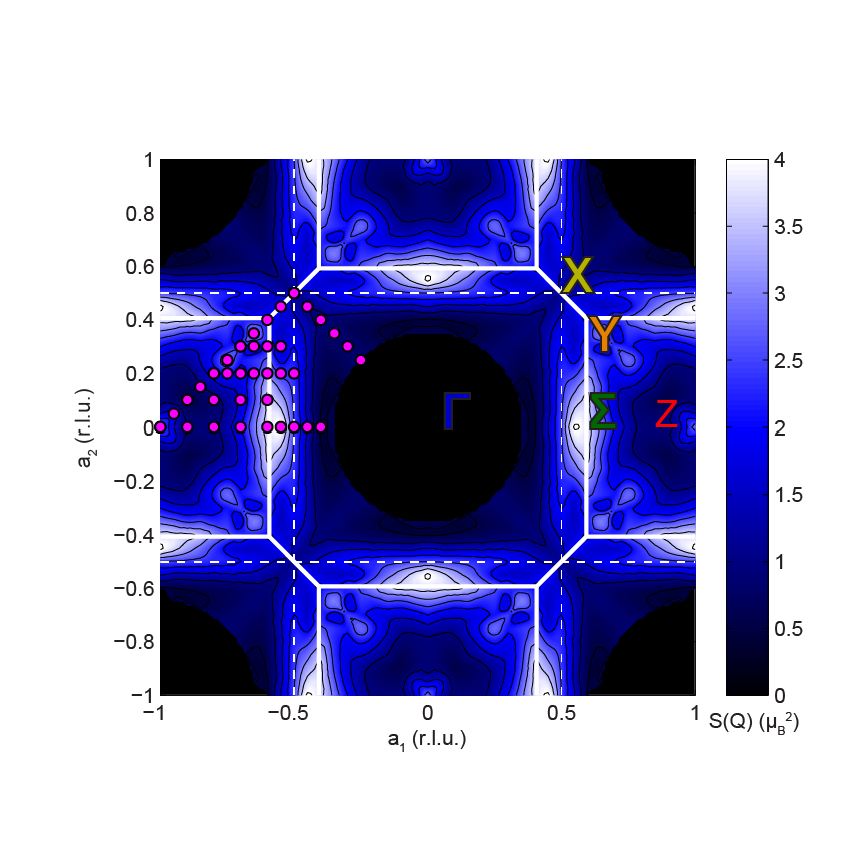}
\end{center}
\caption{Variation of energy-integrated intensity of the magnetic excitations $S(Q)$ in the HO state in the basal plane. Regions of high intensity follow the dispersion minima plotted in Fig.~\ref{2Dmagdisps} and scattering is weak away from the zone boundary. A clear distinction is evident between the excitations near \textbf{Z} and those that follow the BZ edges. Calculation of $S(Q)$ is discussed in the text. Integration of $S(Q)$ over the entire BZ suggests the magnetic excitations in the HO phase are due to a fully-degenerate uranium \emph{f}-state. This plot was symmetrized in the same manner as Fig.~\ref{2Dmagdisps}. Data were taken on BT-7.}
\label{MagI}
\end{figure}

The energy-integrated magnetic scattering intensity $S(Q) = \int S(Q,\omega)d\omega$ in the basal plane is approximated by integrating the area under magnetic peaks measured at constant $Q$ on BT-7, for which we estimate a 30\% absolute uncertainty (Fig.~\ref{MagI}). The calculation first requires determination of the dynamic spin correlation function
\begin{equation}
S(Q,\omega) = \frac{\mu_B^2}{p^{2}|f(Q)|^2 e^{-2W}R}I(Q,E).
\end{equation}
Here $I(Q,E)$ is the measured intensity, $f(Q)$ is the U$^{3+}$ or U$^{4+}$ form factor, which are similar \cite{Kuwahara06}, $\mu_B$ is the Bohr magneton, $e^{-2W} \approx 1$ is the Debye-Waller factor, $p=0.2695 \times 10^{-14}$~m is a proportionality constant for magnetic neutron scattering, and $R$ is a normalization factor determined from acoustic phonon scattering \cite{Xu13}. Figure~\ref{MagI} shows that high intensity scattering traces the local minima plotted in Fig.~\ref{2Dmagdisps}. The intensity is highest near $\pmb{\Sigma}$, diminishes by about $\frac{1}{3}$ approaching \textbf{Y}, and near \textbf{X}, it is already smaller by $\frac{1}{2}$. This variation is consistent with the early results of Broholm et al. \cite{Broholm91}. Again, it is evident that the inelastic magnetic scattering intensity peaks along the zone edges, which also can be seen in the $a-c$ plane in the data of Wiebe and et al. \cite{Wiebe07}.

There is a dramatic difference between $\pmb{\Sigma}$ and \textbf{Z}. Although $S(Q)$ at \textbf{Z} is strong, the intensity clearly decreases by an order of magnitude within a small $q$-window about \textbf{Z} in both the $a-$ and $c-$ directions. This implies that the magnetic excitations at \textbf{Z} have a distinct origin from those at the zone boundary, even if it is possible to draw a continuous dispersion \cite{Janik09}. Indeed, the excitations respond differently to experimental tuning: applied pressure opens the energy gap at $\pmb{\Sigma}$ and closes the gap at \textbf{Z} \cite{Bourdarot05}, and Re substitution selectively suppresses the excitations at \textbf{Z} \cite{Williams12}. These excitations emanate from the AFM zone center and their distinct intensity decrease is reminiscent of the effects of an AFM structure factor.

The extensive determination of $S(Q)$ makes it possible to estimate the effective spin per uranium atom giving rise to the magnetic scattering. This spin is determined from the sum rule
\begin{equation}
\frac{1}{3}(g\mu_B)^{2}J(J+1) = \frac{\int \int_{BZ}S(Q,\omega) d^3Q d\omega}{\int_{BZ}d^3Q},
\end{equation}
where $g$ is the Land\'{e} g-factor, which is 0.73 or 0.8 for U$^{3+}$ or U$^{4+}$, respectively. Taking advantage of the fact that most of the magnetic intensity resides in-plane, combined with our measurements in the $a-c$ plane and the data of Wiebe et al. \cite{Wiebe07}, $S(Q)$ in three dimensions of reciprocal space can be interpolated and integrated over the Brillouin zone. Our calculations yield a range of $J$ values from 4-4.5 that depends on how rapidly the magnetic intensity falls off in the $c-$direction. These $J$ values correspond to full U$^{4+}$ ($f^2, J=4$) or U$^{3+}$ ($f^3, J=\frac{9}{2}$) moments, a fact consistent with the conspicuous absence of direct evidence for splitting of the $J$ multiplet due to crystalline electric fields in URu$_2$Si$_2$. This indicates that all of the magnetic correlations are accounted for, ie, there is no appreciable magnetic spectral weight at higher energy. Also, since the full moment is accounted for in the inelastic channel, there is no spectral weight remaining for elastic scattering, and static long-range magnetic order is inhibited. Completely neglecting the intensity along $c$ and integrating over only the in-plane magnetic intensity yields $J \approx 2$, which still implies a large \emph{f}-state degeneracy that is incompatible with models invoking multipolar order. Because of its limited $q$-range, the inclusion in the integral of the magnetic scattering near \textbf{Z} is of little consequence to the calculated $J$ value, amounting to less than a 5\% correction. This indirectly suggests that the magnetic excitations near \textbf{Z} may be competing with the other excitations, perhaps as an incipient AFM order.

\subsubsection{Temperature dependence}

\begin{figure}
\begin{center}
\includegraphics[width=3.4in]{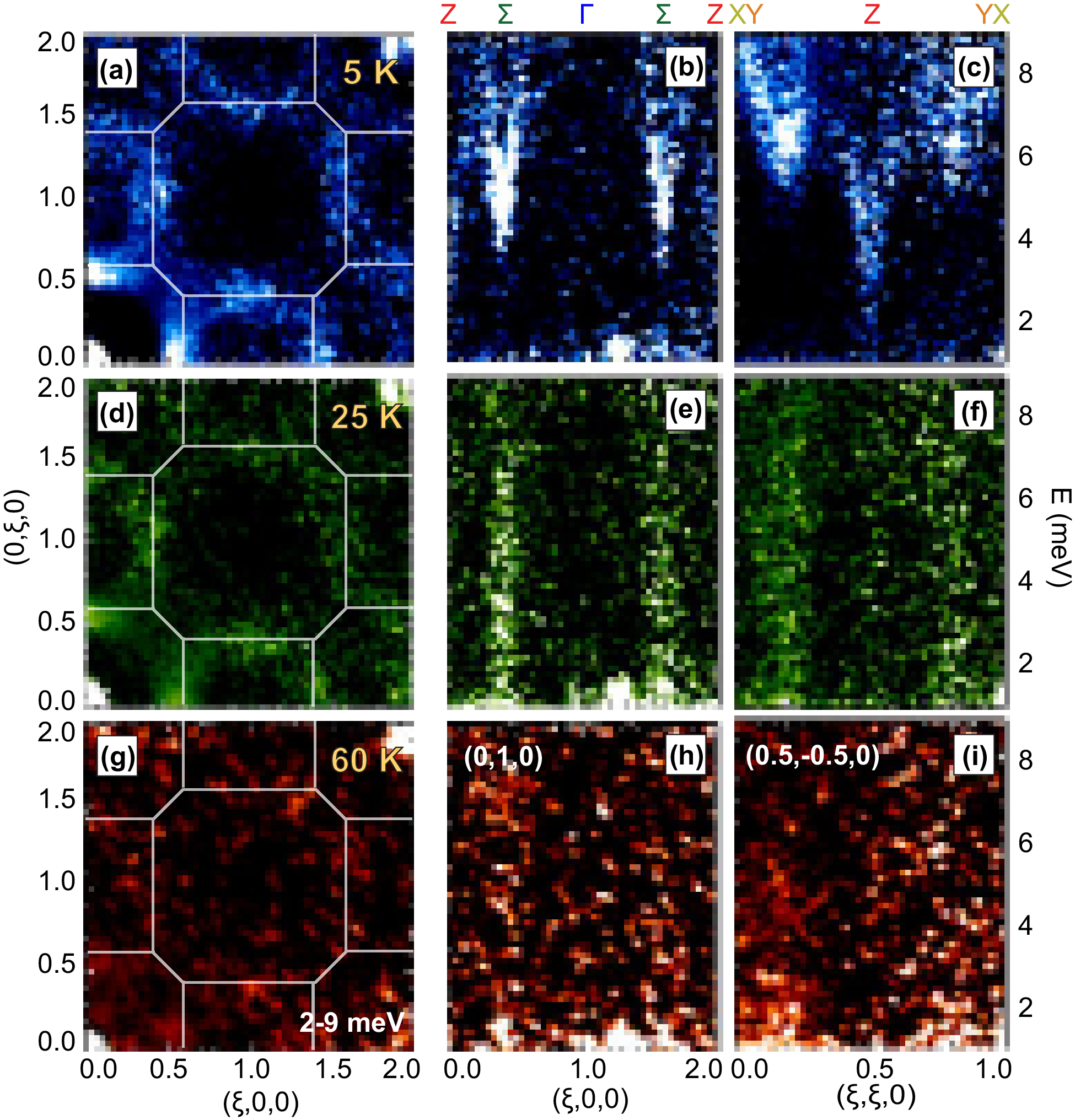}
\end{center}
\caption{Magnetic excitation spectrum above and below $T_\mathrm{HO}$. At 5~K, a) shows the intensity of excitations in the basal plane, integrated from 2-9~meV. In the HO phase, the magnetic excitations are gapped and dispersive (compare to Fig.~\ref{magDOS}), but they respect the high-temperature BCT symmetry, overlaid in white. b,c) Energy dependence of magnetic excitations along selected cuts in the basal plane. The magnetic scattering forms vertical stripes over the 1-9~meV energy range, indicative of overdamping. Specific points in reciprocal space are emphasized in b) $\pmb{\Sigma}$; c) \textbf{Z}, near \textbf{Y} and \textbf{X}. d-f) Same as a-c, but at 25~K. For $T > T_\mathrm{HO}$, the excitations broaden in $Q$ and $E$, but still visibly trace the zone boundaries. g-h) By 60~K, the excitations are no longer strongly peaked near the zone edges. Data were taken on DCS. The intensity scale in Fig.~\ref{DCS} is the same in all panels, and saturates at 3~$\mu_B^2$/U. The constant $Q$ cuts are integrated over a $\pm 0.1$ window.}
\label{DCS}
\end{figure}

At high temperatures, it is established that the magnetic spectrum consists of overdamped modes near the special $q$ points $\pmb{\Sigma}$ and \textbf{Z} \cite{Broholm91,Wiebe07,Janik09,Bourdarot10}. Our new data show that this is true across the entire BZ. This is well demonstrated by time-of-flight inelastic neutron measurements, as plotted in Fig.~\ref{DCS}. First, time-of-flight measurements independently corroborate the $S(Q)$ distribution in the HO phase measured by triple-axis, as is evident by the remarkable agreement between Fig.~\ref{MagI} and Fig.~\ref{DCS}a. The latter differs from the former only in that it lacks $f(Q)$ correction, and the integration is over a fixed $E$ range of 2-9~meV for all $Q$, including all scattering such as that from strong acoustic phonons at the (2,0,0) and (2,2,0) positions. Cuts are taken along $(\xi,0,0)$ at constant $(0,1,0)$ in Fig.~\ref{DCS}b and $(\xi,\xi,0)$ at constant $(0.5,-0.5,0)$ in Fig.~\ref{DCS}c, with the relevant $q$ points indicated above. These plots show magnetic dispersions that are familiar from Fig.~\ref{magDOS}a. Note that Fig.~\ref{DCS}b is consistent with the data of Wiebe et al. \cite{Wiebe07}, while Fig.~\ref{DCS}c shows the qualitative similarity of the \textbf{X}-\textbf{Z}-\textbf{X} cut, in which the distinctness of the \textbf{Z}-centered excitations is even more readily apparent.

Figure~\ref{DCS}d shows the magnetic scattering at 25~K, well above $T_\mathrm{HO}$. Although slightly wider, the integrated intensity still strikingly traces the BCT BZ boundary. Figures~\ref{DCS}e-f show the $E$-dependence along the high-symmetry cuts, in which the magnetic excitations are no longer visibly dispersing, but appear as vertical stripes centered at the $q$ locations of the dispersion minima. This scattering, extending from the elastic line to energy transfers greater than 9~meV, is consistent with overdamped excitations \cite{Broholm87}. Near $\pmb{\Sigma}$, the excitations in the paramagnetic state are actually still dispersive, although the linewidths are extremely broad \cite{Janik09}. Note again the qualitative similarity between the \textbf{Z}-$\pmb{\Sigma}$ and \textbf{Z}-\textbf{X} cuts. In fact, overdamped scattering extends to every $Q$ for which there exists a well-defined magnetic dispersion in the HO phase. The fact that the reciprocal-spatial intensity modulation in Fig.~\ref{DCS}d) exists outside of the HO phase at 25~K establishes it as a characteristic of the highly-correlated paramagnetic high-temperature phase that obeys BCT symmetry. Despite the opening of energy gaps below $T_\mathrm{HO}$, clearly the BCT symmetry of the correlated electrons persists into the ordered phase.\cite{SM} This is a central result of our study.

\begin{figure}
\begin{center}
\includegraphics[width=3.4in]{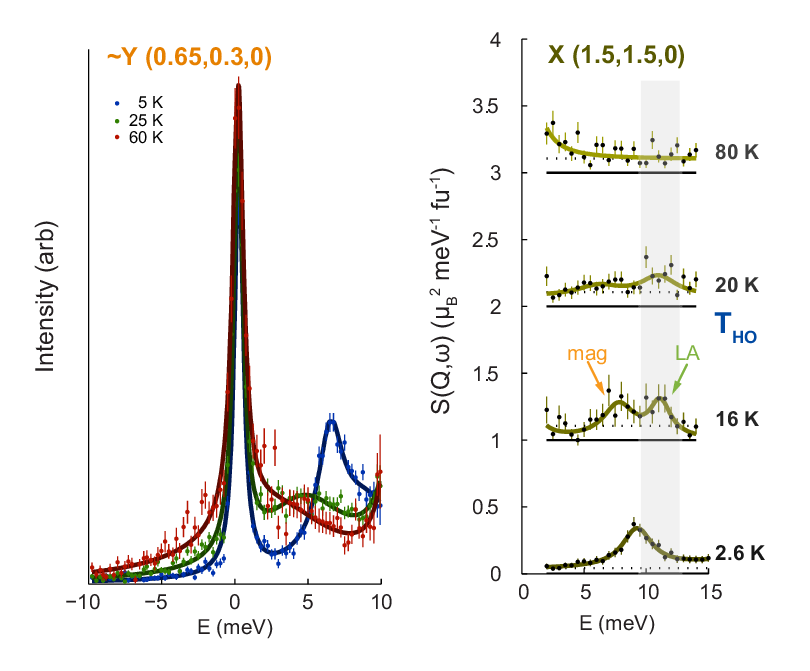}
\end{center}
\caption{Temperature dependence of the magnetic scattering near the \textbf{Y} and at the \textbf{X} points illustrates that high-temperature paramagnetic excitations exist across the BZ and evolve similarly as a function of temperature. At the dispersion minimum near \textbf{Y}, the temperature dependence of the inelastic scattering is similar to that previously reported at $\pmb{\Sigma}$ \cite{Broholm91,Janik09}. The peak energy decreases with increasing temperature and broadens dramatically. At 80~K, the hybridization of \emph{f}-electrons is apparent at the \textbf{X} point where overdamped magnetic excitations mask weak scattering from the LA phonon. Upon cooling, low $E$ excitations develop and narrow into peaks as $T_\mathrm{HO}$ is crossed. The higher-energy excitations are phonons, as indicated. The LA phonon at \textbf{X} is only resolved at intermediate temperatures. \textbf{Y} data were taken on DCS, while \textbf{X} data were taken on BT-7. Phonons are fit by Lorentzian functions.}
\label{YX}
\end{figure}

In the HO phase, there is no magnetic scattering at small $Q$ over 2-9~meV, but as temperature rises, the intensity becomes more evenly distributed. Only at higher temperatures does the $Q$-space variation in intensity reduce, as shown in Fig.~\ref{DCS}g. Figures.~\ref{DCS}h-i also suggest that by 60~K, which is near the coherence temperature, the overdamped excitations are rather homogeneously distributed in $Q$, a characteristic of paramagnetic scattering from local moments. The temperature-dependence of the magnetic scattering at the dispersion minimum near \textbf{Y} is shown clearly in Fig.~\ref{YX}, which also shows a peak at finite energy in the paramagnetic state like the excitations near $\pmb{\Sigma}$ \cite{Broholm91,Janik09}. The peak intensity moves from 6.5~meV at 5~K, to 4.8~meV at 25~K, to 2.2~meV at 60~K, at which temperature a Lorentzian fit has a full width at half maximum of 10~meV. The scattering is highly overdamped and extended across reciprocal space, for $T > T_\mathrm{HO}$. This fact is corroborated at \textbf{X} by BT-7 data, as shown in Fig.~\ref{YX}. Thus the correlations become short-lived and only persist over short lengths as the temperature increases.

One of the most important revelations regarding the nature of the HO phase is that the gapping of the paramagnetic excitations in the vicinity of $\pmb{\Sigma}$ by the HO transition can account for the entropy change at $T_\mathrm{HO}$ \cite{Wiebe07}. Although excitations in the $a$-$c$ plane appear over a limited $q$ range about $\pmb{\Sigma}$, this model does not take into account the magnetic in-plane excitations away from $\pmb{\Sigma}$. Our data show that the excitations do not emanate from a single $q$, but rather that a large extended range of correlations is gapped by the HO transition. This suggests a slight modification of the main result of Wiebe and coworkers, namely, that the DOS of their model \cite{Wiebe07} can be considered as a $q$-integrated approximation of the full magnetic spectrum.

\subsubsection{Relation to electronic structure}

It is unusual that a nominally full uranium moment persists to low temperatures, but this surprising fact yields an important clue to the nature of the high-temperature hybridization between uranium \emph{f}-states and the conduction electrons in URu$_2$Si$_2$. Although often in the literature the coherence temperature of 80~K is equated with a Kondo temperature, the measured basal-plane electrical resistivity has a negative temperature derivative from 80~K up to incredibly high temperatures in excess of 1,200~K \cite{Schoenes87}. This behavior implies a single-ion Kondo temperature $T_K > 350$~K \cite{Schoenes87}, which indicates that the hybridization between local \emph{f}-states and conduction electrons involves an energy scale much larger than 80~K ($\approx 7$~meV). The full moment degeneracy is preserved by the high-temperature Kondo scattering, which serves to effectively quench the local crystal-field splitting. Such an effect is observed in the case of CeNi$_9$Si$_4$, which has a full 6-fold degenerate ground state when chemical tuning sets $T_K = 70$~K, comparable to the crystal-field splitting \cite{Gold12}. The large low-temperature $J$ in URu$_2$Si$_2$ gives rise to a strong magnetic excitation spectrum and, of course, a large configurational degeneracy.

A big challenge in the study of URu$_2$Si$_2$ is that fundamentals of the electronic structure, such as the number of \emph{f}-electrons on the uranium ions, are still hotly debated. There have been many attempts to reconcile first-principles calculations \cite{Denlinger01,Oppeneer10,Elgazzar09,Haule09} with experimental angle-resolved photoemission spectrum (ARPES) \cite{Denlinger01,Santander09,Kawasaki11,Yoshida13,Meng13,Chatterjee13}. A crucial detail is the location of Fermi surface (FS) pockets. Of particular historical concern has been the origin of the incommensurate magnetic excitations at $\pmb{\Sigma}$, for which potential nesting vectors have been proposed \cite{Oppeneer10,Janik09,Balatsky09,Meng13}. However, compared to archetypes such as the charge density wave in uranium \cite{Smith84} and the spin density wave in chromium \cite{Fawcett88}, URu$_2$Si$_2$ lacks a crucial characteristic, namely superlattice reflections, in nuclear, magnetic, or electronic elastic scattering at any known incommensurate or commensurate wavevector. In addition, the broad extent of the magnetic excitations in $Q$ is inconsistent with the magnetic excitations typical of density waves, which have steep dispersions well-localized in $Q$ \cite{Coldea01,Endoh06}, even when the transition is completely suppressed \cite{Fawcett94}.

Various heavy fermion and Kondo insulator systems feature quasielastic paramagnetic scattering, albeit with widely-varying material-dependent reciprocal-space structure, and energy and temperature scales \cite{Holland82,Aligia85,Broholm87b,Rossat88,Sato95,Christianson06,Hiess06}. More generally, calculations show that the intensity of the magnetic scattering due to a hybridized band structure is strongest at the BZ boundaries \cite{Aligia85,Brandow88}, which has been confirmed in some intermediate valent materials \cite{Aligia85,Christianson06,Alekseev07,Fanelli14}. At the heart of the effect are strong interband transitions across an indirect hybridization gap, from the valence band maximum at the BZ edge to the conduction band minimum located at the BZ center of, for example, the archetypal hybridized bands \cite{Brandow88}
\begin{equation}
E_{\pm} = \frac{1}{2}\{E_k+E_f \pm [(E_k-E_f)^2+4V_k^2]^\frac{1}{2}\},
\end{equation}
where $E_k$ and $E_f$ are the itinerant and \emph{f}-state dispersions, and $V_k$ is the hybridization potential. This type of scattering has three important features: it yields both quasielastic and inelastic $E$-asymmetric lineshapes, it persists even if the gap is not full, or does not sit at the chemical potential, and the intensity is strongest along the zone boundaries \cite{Aligia85,Brandow88}. The similarities with the magnetic excitations of URu$_2$Si$_2$ make this framework a good starting point for exploring specific details of the electronic structure.

The electronic structure of URu$_2$Si$_2$ near \textbf{Z} has been studied extensively by ARPES \cite{Ito99,Meng13,Boariu13,Yoshida13,Chatterjee13,Denlinger01,Kawasaki11,Yoshida12}. A small hole pocket is centered on \textbf{Z} with an in-plane Fermi wavevector $k_F \approx 0.2$~\AA$^{-1} \approx 0.15$~r.l.u. A narrow band with \emph{f}-character sharpens inside this pocket \cite{Chatterjee13}, consistent with expectations of \emph{d}-\emph{f} hybridization in a periodic Anderson model \cite{Denlinger01}. A second, larger hole-like pocket is also centered on \textbf{Z} \cite{Kawasaki11,Denlinger01} with $k_F \approx 0.4$~r.l.u. along $\pmb{\Gamma}$-\textbf{Z}. Remarkably, the extent of this pocket is nearly identical to the ring of inelastic magnetic scattering (Figs.~\ref{2Dmagdisps},\ref{DCS}). This coincidence is an important clue to the origin of the magnetic excitations as well as the underlying FS.

Essentially, the two \textbf{Z}-centered pockets can account for the bulk transport properties of URu$_2$Si$_2$. The small hole pocket matches the dimensions of the $\alpha$ pocket measured by quantum oscillations \cite{Hassinger10} with an approximate carrier density of $2 \times 10^{20}$~cm$^{-3}$, which agrees well with the effective Hall carrier density in the HO phase \cite{Kasahara07}. The bigger hole pocket, which is not observed in quantum oscillations \cite{Ohkuni99,Nakashima03,Jo07,Hassinger10,Altarawneh11}, has a carrier density of order $10^{21}$~cm$^{-3}$, which agrees well with the larger Hall carrier density at high temperatures \cite{Schoenes87}. The carrier density decrease and absence of this larger pocket in the HO phase suggests that the large \textbf{Z} FS is gapped at low temperatures, whereas the experimental similarity of the quantum oscillations, bulk properties, and point contact spectroscopy \cite{Lu12} between HO and AFM phases implies that it is gapped in both HO and AFM states. A charge gap is detected at $T>T_\mathrm{HO}$ \cite{Levallois11,Park12,Shirer13}, which suggests that it is not directly responsible for the entropy change at $T_\mathrm{HO}$.

\begin{figure}
\begin{center}
\includegraphics[width=3.4in]{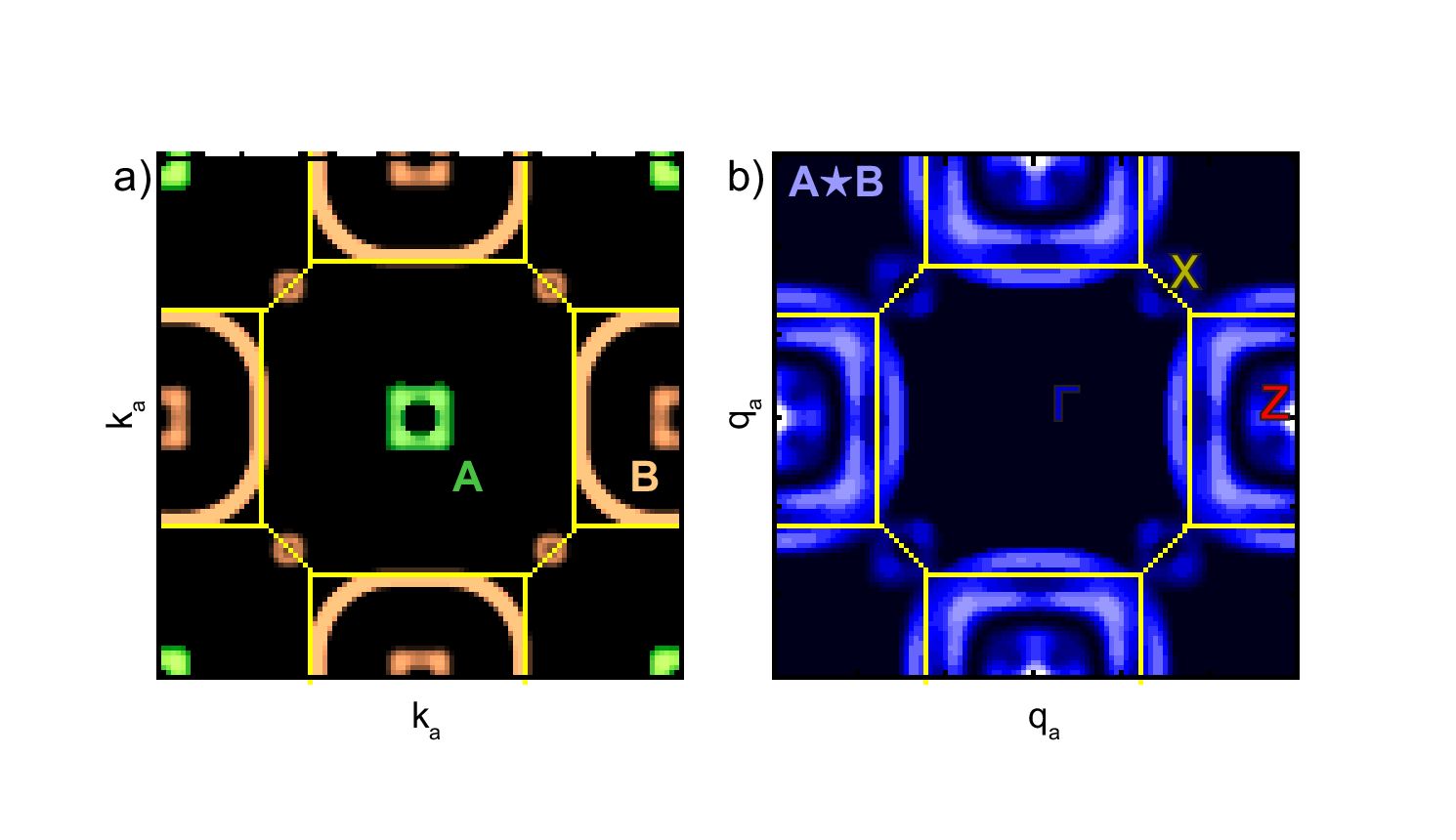}
\end{center}
\caption{Calculation of interband scattering due to Fermi surface inferred from ARPES and quantum oscillations measurements. a) A model Fermi surface consists of an electron pocket at $\pmb{\Gamma}$ (A) and hole pockets centered on \textbf{Z} and \textbf{X} (B). b) Calculated cross-correlation A$\star$B shows the associated scattering vectors and relative intensities, whose agreement with our experimental data in Figs.~\ref{MagI} and \ref{DCS} suggests that the magnetic excitations are due to interband scattering.}
\label{ARPES}
\end{figure}

Figure~\ref{ARPES}a shows a model FS inferred from the previous discussion that consists of an electron pocket (green) at $\pmb{\Gamma}$ and several hole pockets (orange). The cross-correlation of the two, Fig.~\ref{ARPES}b, which maps the $q$ vectors of possible interband transitions, strongly resembles the magnetic scattering intensity maps (Figs.~\ref{MagI} and \ref{DCS}). Cross-correlation can be used to identify nesting vectors between Fermi surfaces involved in density-wave formation \cite{Borisenko08}, but in the case of magnetic scattering in URu$_2$Si$_2$, the excitations span a range of $E$. Considering that the general idea of interband excitations holds also for Kondo insulators, the relevant pockets need not actually sit at the chemical potential. As with the simple hybridized bands $E_{\pm}$, the essential ingredients behind extensive zone edge excitations are a hole-like band at the BZ edge, and an electron-like band at the BZ center. The first condition is satisfied by the large \emph{Z}-centered FS, while the second condition implies the existence of a small electron pocket at $\pmb{\Gamma}$. In addition, the small \textbf{Z} pocket is responsible for the AFM-like excitations.

The existence of an electron pocket at $\pmb{\Gamma}$ is suggested by first-principles calculations \cite{Denlinger01,Elgazzar09} that are supported by ARPES measurements \cite{Meng13}. However, other ARPES studies conclude that it is actually hole-like, with $k_F \approx 0.2$~{\AA}$^{-1}$ and an unusual dispersion \cite{Yoshida13,Boariu13,Chatterjee13}, and it is not detected using higher incident energies \cite{Kawasaki11}. Although this FS pocket requires further investigation, the evidence suggests that hybridization plays an important role in the vicinity of $\pmb{\Gamma}$. Further, note that the inelastic scattering does not necessarily require that relevant bands cross the chemical potential.

In order to account for the magnetic scattering near \textbf{X}, our model predicts that there is an additional small hole pocket there (Fig.~\ref{ARPES}). As with the pocket at $\pmb{\Gamma}$, there is some experimental uncertainty regarding its properties. The \textbf{X} point was originally found to harbor well-defined small square hole-like pockets \cite{Denlinger01} that have since been attributed to surface states \cite{Denlinger11}. More recently, Meng et al. have shown evidence for a small X pocket that demonstrates broken lattice symmetry \cite{Meng13}, but soft x-ray ARPES is not sensitive to this pocket either \cite{Kawasaki11}. We tentatively place the small $\beta$ pocket detected by quantum oscillations \cite{Ohkuni99,Hassinger10} at \textbf{X}. With an effective mass of 25 times the bare electron mass, the band that makes up the $\beta$ pocket is very flat and within 1~meV of the Fermi level, making detection via ARPES difficult. In general, ARPES studies suggest that the identified FS pockets all coexist with a narrow \emph{f}-state that sits just below the Fermi level \cite{Chatterjee13,Boariu13}, which is consistent with the fractional \emph{f}-count determined by electron energy loss spectroscopy \cite{Jeffries10}.

\subsection{Phonons}

\subsubsection{General characteristics}

\begin{figure}
\begin{center}
\includegraphics[width=3.4in]{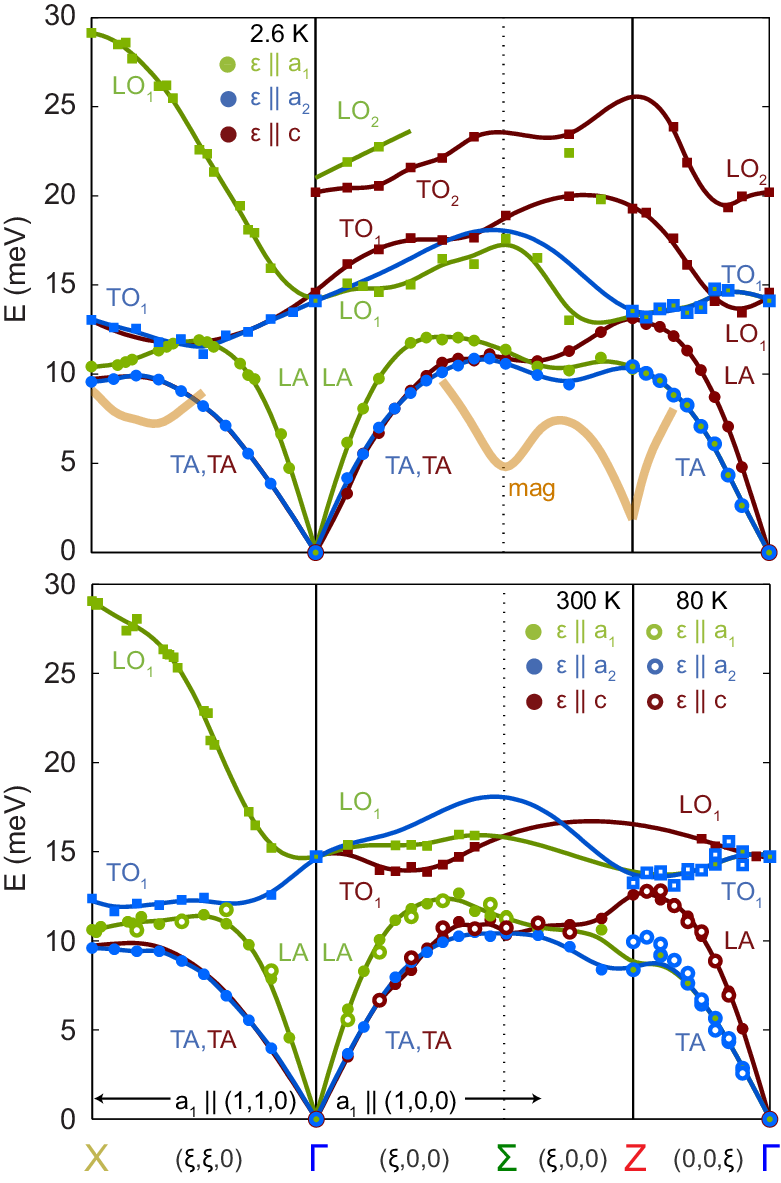}
\end{center}
\caption{Phonon dispersions at different temperatures. The high-temperature dispersions are largely similar to those in the HO phase. However, there are some notable exceptions. A dramatic difference in the 2.6~K data is that the low-lying $c$-polarized TO$_1$ mode lies at lower $E$ than the LO$_1$ mode along $\pmb{\Gamma}$-$\pmb{\Sigma}$ and lacks the local minimum near $\pmb{\Gamma}$ along the $\pmb{\Gamma}$-\textbf{Z} direction. Most features were determined via inelastic neutron scattering on BT-7, but the LO$_1$ $\pmb{\Gamma}$-\textbf{X} mode was determined at 2.0 GPa using inelastic x-ray scattering on HERIX. Colors denote phonon polarization, while shapes delineate acoustic and optic modes. Magnetic excitations (thick lines) are included for reference.}
\label{hiTdisps}
\end{figure}

The crystal lattice plays a seemingly passive role, showing no signs of broken symmetry in the HO phase \cite{Kernavanois99}. Meanwhile, the excitation spectrum of the crystal lattice is poorly characterized beyond small $Q$. The energies of Raman-active optic phonons show minimal temperature dependence but signs of low-temperature electron-phonon coupling \cite{Cooper87,Lampakis06}. Ultrasound studies, which probe the very low-$E$ acoustic phonons, show tendencies toward symmetry-breaking: softening is observed in a volume-conserving, symmetry-breaking mode below 70~K \cite{Kuwahara97}, and this softening disappears when high field destabilizes the HO phase \cite{Yanagisawa13}. There is also an increase in thermal conductivity at $T_\mathrm{HO}$ \cite{Behnia05} that has been argued to arise from the electrostatic coupling of the HO parameter to the lattice \cite{Sharma06}. Such coupling suggests that the phonons might display the in-plane magnetic and electronic anisotropy inferred from recent torque magnetometry \cite{Okazaki11} and cyclotron resonance measurements \cite{Tonegawa12}.

\begin{figure}
\begin{center}
\includegraphics[width=3.4in]{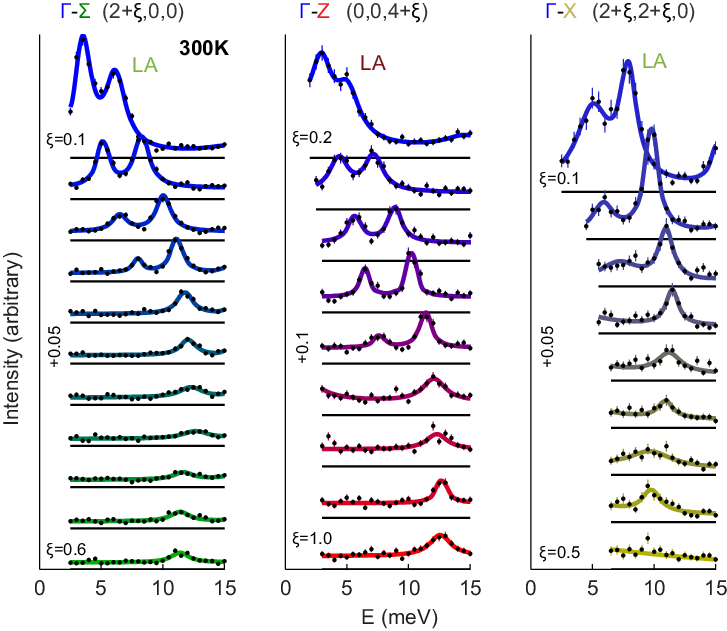}
\end{center}
\caption{Example phonon data from BT-7 at 300~K. These constant $Q$ scans measured along high-symmetry directions are sensitive to longitudinally polarized phonons. The data mostly cover the energy range of the LA modes, although LO$_1$ mode is seen at small $\xi$ along $\pmb{\Gamma}$-\textbf{X}. These modes are summarized in Fig.~\ref{hiTdisps}. Lines are fits to Lorentzian functions and constant background. Prominent low-energy peaks at small $\xi$ are due to intense, nominally-forbidden, TA phonons that appear due to instrumental resolution. Along $\pmb{\Gamma}$-\textbf{X}, the data for $\xi > 0.2$ correspond to symmetrically-equivalent points at $\xi < -0.2$, towards $Q = (1.5,1.5,0)$.}
\label{RTphonon}
\end{figure}

\begin{figure}
\begin{center}
\includegraphics[width=3.4in]{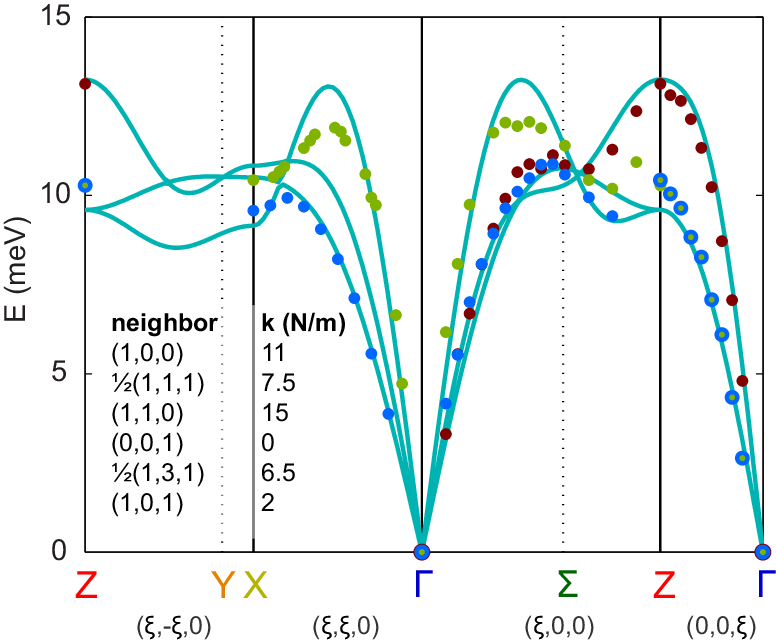}
\end{center}
\caption{One-atom Born von Karman force constant model that approximates the acoustic phonon spectrum. The force constants corresponding to the selected 5 nearest and next-nearest neighbor atoms are listed. The last two force constants help to reduce the LA-TA energy difference, which is otherwise set by the $a/c$ ratio.}
\label{BvK}
\end{figure}

An evaluation of the low-energy phonon dispersions (Fig.~\ref{hiTdisps}) yields no signs of broken symmetry. Example room-temperature data are shown in Fig.~\ref{RTphonon}. Phonons propagating in the basal-plane can have longitudinal and transverse in-plane polarizations, denoted $a_1$ and $a_2$. For phonons propagating along the $c$-axis, the two transverse $a$ polarizations are degenerate by tetragonal symmetry: note the transverse acoustic TA and transverse optic TO$_1$ phonons. Any lifting of this degeneracy due to dynamic symmetry breaking, say towards an orthorhombic distortion, is not observed. Born-von Karman force-constant modeling can reproduce the acoustic modes using a simplified BCT crystal consisting of only one uranium atom and 5 force constants (Fig.~\ref{BvK}). Next-nearest-neighbor interactions are a proxy for the effects of the Ru and Si atoms, but the agreement indicates that the acoustic phonons are well-behaved. Based upon the model we conservatively interpolate the \textbf{Z}-\textbf{X} acoustic modes, which were not experimentally observed. These zone edge modes have flat dispersions with energies greater than the measured magnetic excitations (Fig.~\ref{magDOS}), lending additional confidence in our integration of the magnetic intensity (Fig.~\ref{MagI}).

There is no obvious anomalous $q$-dependence in the measured acoustic modes in the vicinity of the magnetic excitations that might suggest strong magneto-elastic coupling. Inelastic x-ray scattering measurements confirm that the LA phonon along $\pmb{\Gamma}$-\textbf{X} is well-defined, complementing polarized neutron scattering measurements to help distinguish it from the magnetic scattering at the same $q$ and $E$ (Fig.~\ref{HOmag}). This means that the appearance of a weak phonon peak at high temperatures (Fig.~\ref{YX}) is due to the significant overlap of strong paramagnetic scattering. Furthermore, the softening observed in ultrasound studies \cite{Kuwahara97,Yanagisawa13} appears to be limited to very small $Q$ and does not produce any dramatic effects at higher $Q$ in the acoustic phonons. In contrast, the optic phonons do show some unusual behavior.

\begin{figure}
\begin{center}
\includegraphics[width=3.4in]{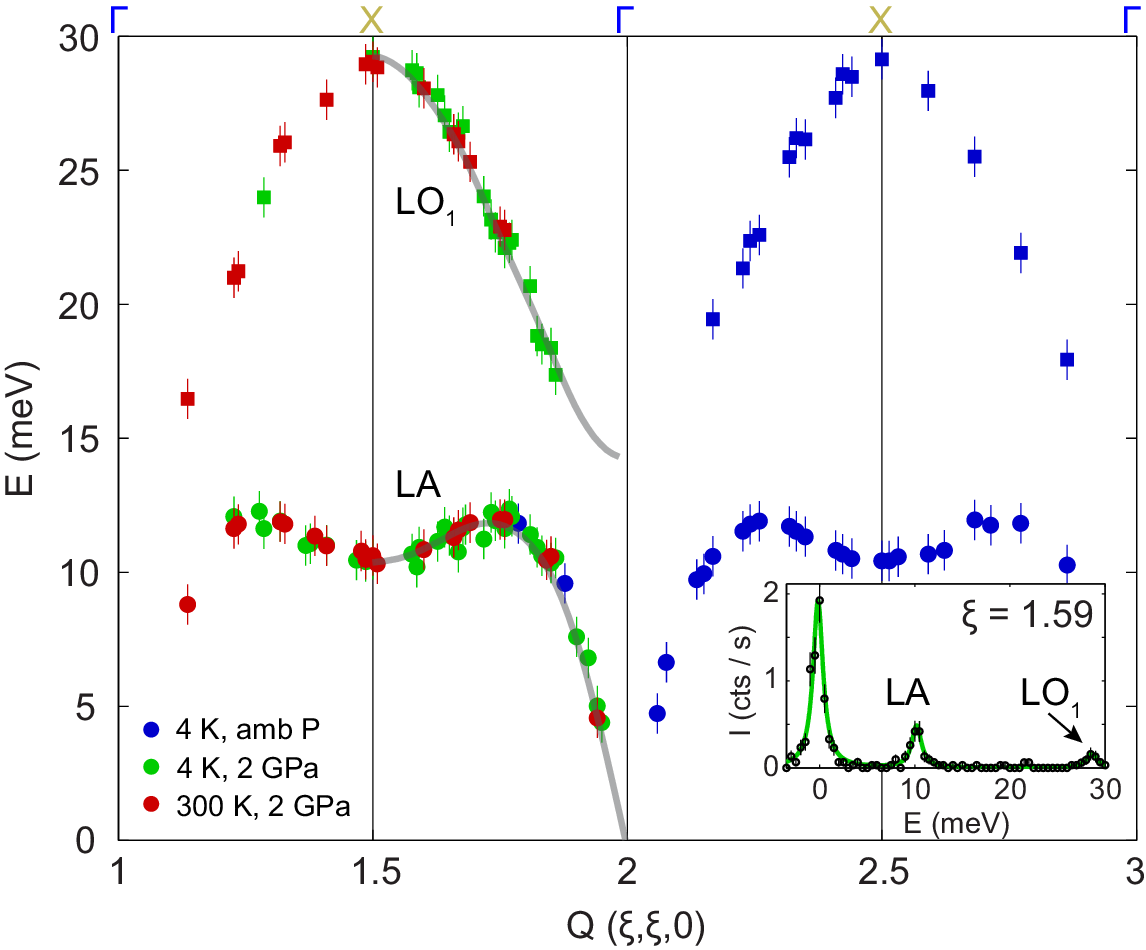}
\end{center}
\caption{Phonon dispersions determined using inelastic x-ray scattering on HERIX clearly show the LA and LO$_1$ phonons along the $\pmb{\Gamma}$-\textbf{X} direction through several Brillouin zones. These dispersions are roughly independent of temperature and do not change with the modest applied pressure of 2.0~GPa that induces long-range AFM order. Error bars correspond to the energy resolution. Lines are guides to the eye. The inset shows an example scan at constant $Q$, taken at 4~K and 2~GPa.}
\label{HERIX}
\end{figure}

The optic modes detected by neutrons match those observed by infrared and Raman spectroscopies at small $Q$. The lowest-energy optic O$_1$ phonons intersect the zone center $\pmb{\Gamma}$ at 14~meV, corresponding to the $a$-polarized infrared-active mode \cite{Levallois11}. The measured O$_2$ phonons include the Raman-active 20~meV excitation with B$_{1g}$ symmetry \cite{Cooper87}. Of particular note is the huge in-plane anisotropy of longitudinally-polarized optic LO$_1$ modes, which is readily apparent via a comparison of the two high-symmetry basal-plane directions $\pmb{\Gamma}$-$\pmb{\Sigma}$ and $\pmb{\Gamma}$-\textbf{X} that are rotated with respect to each other by $45^{\circ}$ (Fig.~\ref{hiTdisps}, in green). The upward-dispersing $\pmb{\Gamma}$-\textbf{X} branch was originally difficult to identify in neutron scattering measurements, and was confirmed across several BZs via inelastic x-ray scattering (Fig.~\ref{HERIX}) that is sensitive to only to the phonon excitations. Although it disperses strongly, the LO$_1$ mode lacks any notable temperature or pressure dependence, indicating that it does not play an important role in the development of electronic correlations or the ordered phases. Its presence may be understood by analogy to a similar steep dispersion observed in UO$_2$ \cite{Pang13}, which suggests that the LO$_1$ dispersion results from the large mass differences of the constituent Si, Ru, and U atoms. This distribution is also responsible for the large $E$ range of the phonon spectrum. At higher energies, other known phonon modes at $\pmb{\Gamma}$ are IR-active at 42.4~meV ($a$-polarized) and 47~meV ($c$-polarized) \cite{Levallois11} and Raman-active at 55~meV (A$_{1g}$ $c$-polarized) \cite{Cooper87}.

\subsubsection{Temperature dependence}

The phonon temperature dependence highlights important electronic interactions, particulary among the $c$-polarized phonons. First, note that near \textbf{Z}, LO phonon softening along $\pmb{\Gamma}$-\textbf{Z} is absent at 300~K, in contrast to the typical high-temperature phonon softening that arises due to thermal expansion. In the HO phase, in both the $c$-polarized LO$_1$ and LO$_2$ modes along $\pmb{\Gamma}$-\textbf{Z}, there is a minimum in the dispersion 20-30$\%$ away from $\pmb{\Gamma}$. These features contrast with the relatively flat TO$_1$ dispersion at the same temperature. This low-temperature phonon softening at localized $q$ is a sign of electronic interactions and it is compelling that at least 2 optic modes having different symmetries are affected in a similar manner. We surmise that this effect may be related to hybridization of the FS pocket near $\pmb{\Gamma}$ \cite{Meng13,Yoshida13,Boariu13}.

Looking in-plane, the LO$_1$ modes along $\pmb{\Gamma}$-$\pmb{\Sigma}$ exhibit an extreme temperature dependence. The TO$_1$ $c$-polarized mode is much softer at 300~K than at 2.6~K (Fig.~\ref{hiTdisps}), which results in the inversion of the TO$_1$ mode with respect to the LO$_1$ mode along $\pmb{\Gamma}$-$\pmb{\Sigma}$. This unusual switch is a sign of strong temperature-dependent electron-phonon interactions along the $\pmb{\Gamma}$-$\pmb{\Sigma}$ direction. Yet the fact that strong softening occurs over such a broad range of $q$ is intriguing, especially since our FS model (Fig.~\ref{ARPES}) has no nearby pockets. Another case of softening with increasing temperature is evident near the \textbf{Z} point, which occurs in the $a$-polarized TA phonons.  At 300~K, these show an $E$ drop before the zone edge is reached as well as along the zone boundary $\pmb{\Sigma}$-\textbf{Z}. By 80~K, the dispersion is similar to that observed in the HO phase. The measured difference of 2~meV represents a large 20~\% change. The expected relative change in $E$ is conventionally related to the relative change in lattice volume V by $\frac{\partial E}{E} = -\gamma \frac{\partial V}{V}$, where $\gamma$ is a Gr\"{u}neisen parameter with a value of approximately 2. Between 300~K and 2.6~K, $\frac{\partial V}{V} \approx 10^{-3}$, from which we would expect a change smaller by two orders of magnitude. This large change hints that there are important changes in electronic structure near \textbf{Z} already occurring between 80~K and 300~K, likely associated with the small \textbf{Z} pocket (Fig.~\ref{ARPES} inferred from the limits of the excitations centered on \textbf{Z} in the HO phase (Figs.~\ref{MagI},\ref{DCS}).

\subsection{Thermodynamics}

\begin{figure}
\begin{center}
\includegraphics[width=3.4in]{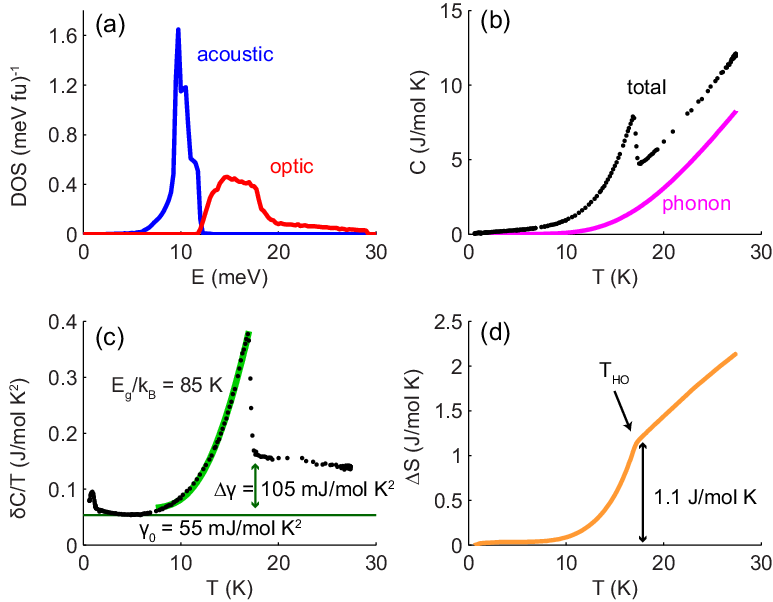}
\end{center}
\caption{Thermodynamic quantities derived from phonon dispersions. a) Density of states (DOS) for acoustic and lowest band of optic modes. b) Comparison of total experimental specific heat \cite{Butch10b} and calculated phonon contribution. c) Difference of curves in b yields effective electronic and magnetic contributions to specific heat, shown here divided by temperature. The contribution in the paramagnetic phase actually decreases with increasing temperature. d) Calculated entropy in excess of the low-temperature $\gamma_0$ electronic contribution. The HO transition liberates 1.1~J/mol~K, or 20$\%$ of $R\ln2$.}
\label{DOS}
\end{figure}

Knowledge of the thermodynamic properties of the electronic and magnetic states in URu$_2$Si$_2$ is fundamental to their proper characterization. Yet the absence of a quantitative treatment of the phonons \cite{Park02} has left a gap in our understanding because the phonons contribute significantly to the measured specific heat \cite{Maple86,Palstra85}. Our measurement of the low-$E$ phonon dispersions makes possible the most accurate experimental determination of the phonon contribution to the specific heat at low temperatures, and yields a confident subtraction from the experimentally measured specific heat. The phonon DOS was calculated by building a histogram of the interpolated 3-dimensional phonon dispersions over the entire BZ. The result of this analysis is shown in Fig.~\ref{DOS}. In Fig.~\ref{DOS}a, the \emph{model-independent} partial DOS of the acoustic (LA,TA) and optic (LO$_1$,TO$_1$) phonons is shown, from which the phonon specific heat curve in Fig.~\ref{DOS}b (magenta) is calculated. The measured specific heat data are taken from Reference~\cite{Butch10b}. Figure~\ref{DOS}c shows the subtracted specific heat $\delta C$, divided by temperature, which is the electronic/magnetic contribution that accounts for the entropy change due to the HO transition. Calculated values for an effective electronic specific heat coefficient $\gamma_0 = 55$~mJ mol/K$^2$ in the HO phase and 160~mJ mol/K$^2$ above the transition. These values can be interpreted as indicating a large reduction of the electronic DOS at the Fermi level \cite{Maple86}, or a removal of magnetic states \cite{Wiebe07}. The anomaly associated with the HO transition is well described by a form $\delta C \propto \exp{(\frac{-E_\mathrm{g}}{k_\mathrm{B} T})}$, where $E_\mathrm{g}/k_\mathrm{B} =  85$~K. This feature is often related to the opening of an energy gap in the electronic DOS. The energy scale corresponds to that of the coherence temperature, or about $5 \times T_\mathrm{HO}$. A calculation of the entropy released by the transition, $\Delta S = \int dT(\frac{\delta C}{T}- \gamma_0)$, yields 1.1~J/mol~K $=0.13$~$k_\mathrm{B}$/fu (there is 1 uranium atom per formula unit), or only 20$\%$ of $R\ln2$ expected from the lifting of degeneracy of a localized electronic doublet, which indicates that itinerant states are responsible for the transition. These values are in general agreement with early results \cite{Palstra85,Maple86}.

The phonon DOS deduced from measurements further uncovers some interesting features in the temperature-dependence of the specific heat. The typical Debye approximation to the phonon specific heat \cite{Schlabitz86} is inaccurate here because there are significant deviations of the measured phonon DOS from the Debye model at low $E$, and there is no reliable temperature range over which to fit the low-temperature approximate form of $C = \gamma T + \beta T^3$. There is particular uncertainty regarding the temperatures above the HO transition, where the $T^3$ approximation is well outside its range of applicability. The properly subtracted specific heat yields a slightly sublinear specific heat, seen as $C(T)/T$ having a negative slope, above the HO transition, as indicated in Fig.~\ref{DOS}c. This behavior can be understood as the result of a decreasing electronic specific heat coefficient due to reduced \emph{f}-electron hybridization as temperature increases. This interpretation is consistent with observation of other hybridization-related phenomena in this temperature range, in particular the closing of spectroscopic gaps \cite{Levallois11,Park12} above the HO transition temperature. In particular, there is no evidence of a maximum that could be ascribed to crystal-field-split local \emph{f}-states. It is also noteworthy that below 5~K, but above the superconducting transition, $C(T)/T$ also has a negative slope. This may be due to the effect of exotic superconducting fluctuations \cite{Yamashita14}.

\subsection{Discussion}

Our data show that inside the HO phase, neither the lattice nor magnetic excitations obey the ST symmetry of the AFM phase. Instead, the gapped magnetic excitations in the HO phase (Fig.~\ref{2Dmagdisps}) follow the same general $Q$-dependence observed by the overdamped modes at high temperature in the BCT phase (Fig.~\ref{DCS}). The behavior in both phases is consistent with interband scattering in the context of a hybridized electronic structure, which involves no spatial symmetry breaking. This indicates that the spatial symmetry of the electronic structure does not change significantly through $T_\mathrm{HO}$, despite the opening of a magnetic gap with its concomitant entropy change.

\begin{figure}
\begin{center}
\includegraphics[width=3.4in]{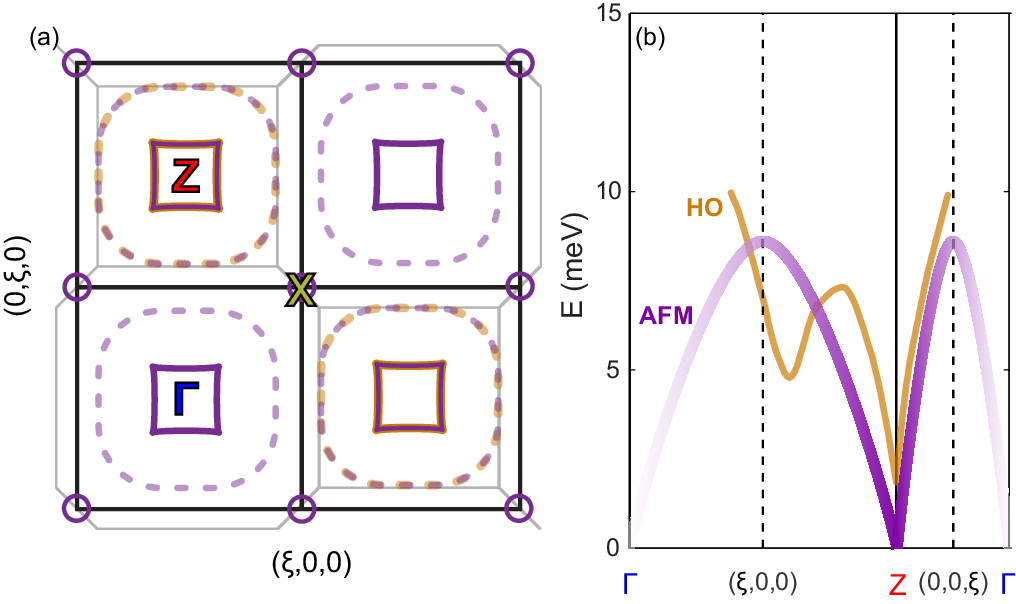}
\end{center}
\caption{Schematic comparison of HO/BCT and AFM/ST symmetries. a) Comparison of the FS pockets in the BCT (orange) and ST (purple) lattices in the basal plane. In the latter, the $\pmb{\Gamma}$ and \textbf{Z} points become equivalent, and the BZ boundary changes from light gray to black. Considering the FS pocket assignments described in the text (see Fig.~\ref{ARPES}), the zone folding does \emph{not} change the extremal cross-sections. The dashed FS pocket is gapped in both the HO and AFM phases. b) The magnetic excitations in the HO phase (orange) do not observe the mirror symmetry required by the ST BZ boundaries (vertical dashed lines), whereas the AFM excitations (purple) should; these curves are based upon limited data under pressure \cite{Villaume08,Bourdarot10}.}
\label{AFM}
\end{figure}

In the absence of long-range AFM order, there are two experimental justifications typically offered in favor of electronic ST symmetry in the HO phase. First, quantum oscillation frequencies do not change discontinuously as pressure tunes the transition between HO and AFM phases \cite{Hassinger10}. Yet, the indifference of the quantum oscillations is also consistent with our inferred FS (Fig.~\ref{ARPES}). Since zone-folding does not intersect any of the pockets and thus does not change their cross-sectional areas, as shown in Fig.~\ref{AFM}a, the quantum oscillations remain unchanged through the pressure-induced BCT-ST (HO-AFM) transition. However, while quantum oscillations need not be sensitive to the symmetry change, magnetic dispersions arising from interband scattering should be strongly dependent upon symmetry, and we do not observe any change in their symmetry. Second, ARPES data show evidence for the identity of $\pmb{\Gamma}$ and \textbf{Z} \cite{Boariu13,Yoshida13}, or two overlapping pockets at \textbf{X} \cite{Meng13}, implying BZ folding below $T_\mathrm{HO}$. These particular features could also arise from BCT-ST folding of our inferred FS, but definitive interpretation is complicated by disagreements between ARPES reports, and possible contributions from surface states as well as minority bulk AFM phases.

Both the HO and the correlated paramagnetic state involve coupling between electron and hole FS pockets. It is therefore natural to consider order parameters composed of the same ingredients. Itinerant, anisotropic order parameters based on particle-hole pairing with a commensurate ordering vector have been suggested and discussed extensively for URu$_2$Si$_2$ \cite{Ramirez92,Tripathi05}. Generally, these have the desirable properties of producing a specific heat anomaly associated with a gapped electronic density of states, as well as the absence of a static moment \cite{Ikeda98} such that they are difficult to experimentally measure. There are numerous related versions, including orbital order \cite{Tripathi05} and spin nematic order \cite{Fujimoto11}, as well as some novel proposals such as spin-orbit order \cite{Das12}. Although certain proposals have been excluded by experiment \cite{Wiebe04}, it would be useful to revisit ideas such as the d-wave SDW \cite{Ikeda98} using our model band structure to calculate the dynamic susceptibility with a focus on its $Q$-dependence. We emphasize that the limited $Q$-range of excitations typical of density-wave orders \cite{Janik09} is incompatible with the extended $Q$-range of the magnetic excitations in URu$_2$Si$_2$. This may be reconciled through a combination of paramagnetic correlated interband scattering and an unconventional density wave order that couples only the small $\pmb{\Gamma}$ and \textbf{Z} pockets. For this scenario, important questions to address include how a density wave onset increases correlation lifetimes across $Q$ such the excitations become sharp, and why associated BCT-ST folding does not introduce magnetic scattering at new $Q$.

Another candidate symmetry reduction currently being debated is electronic orthorhombic distortion \cite{Okazaki11,Kambe13,Tonegawa12,Fujimoto11,Ikeda12}. In the high-temperature paramagnetic phase, an underlying symmetry-breaking tendency is inferred from softening in the $\Gamma_3$ ultrasonic mode below the coherence temperature \cite{Kuwahara97}, which disappears along with the HO phase in applied high magnetic fields \cite{Yanagisawa13}. We observe no signatures of this in the phonon dispersions (Fig.~\ref{hiTdisps}), which means that such effects are limited to the long wavelengths and low energies probed by ultrasound experiments. In principle, it is impossible to completely exclude a tiny nematic order parameter, although it is doubtful that it could account properly for all of the dynamic susceptibility changes as well as the entropy.

Finally, we note several reasons to instead consider $Q=0$ ordering, which is an acknowledged competing ground state of unconventional density waves \cite{Ikeda99}. First, $Q=0$ phases that are not ferromagnets are generally difficult to detect experimentally. Second, there are incipient $Q=0$ correlations in URu$_2$Si$_2$ that maintain strongly correlated behavior \cite{Bauer05,Krishnamurthy08,Butch09}, as evidenced by proximity of the HO phase to a chemically-tuned ferromagnetic (FM) instability \cite{Dalichaouch89,Dalichaouch90,Jeffries07,Butch10b} and the presence of a minority-volume FM phase in some samples \cite{Uemura05}. Third, a $Q=0$ order parameter has no associated change in electronic structure. Finally, the sharp first-order phase boundary that separates the HO and pressure-induced AFM phases \cite{Amitsuka03,Amato04,Amitsuka07,Butch10} is naturally understood as the consequence of different spatial symmetry. The symmetry difference should also be reflected in the magnetic dispersion in the AFM phase: in the simple type A AFM structure, magnetic excitations emanate from the identical magnetic reciprocal lattice point at \textbf{Z} and $\pmb{\Gamma}$, reaching a maximum energy at (0.5,0,0)(Fig.~\ref{AFM}b), and if the minimum at $\pmb{\Sigma}$ survives, a mirror-symmetric copy should be observed in the ferromagnetic zone. This scenario is nominally consistent with the experimental result that under pressure, the excitation at $\pmb{\Sigma}$ increases to 8~meV, while the gap at \textbf{Z} closes \cite{Villaume08,Bourdarot10}.

To summarize, in both the paramagnetic and HO phases, the electronic structure has the same $Q$-dependence. However, the inferred FS agrees in detail with neither calculations that treat the \emph{f}-electrons as primarily itinerant \cite{Elgazzar09,Oppeneer10} nor as localized \cite{Haule09}. Although the magnetic excitations are of itinerant origin, the integrated susceptibility yields a full uranium $J$, which implies that the full \emph{f}-state manifold is hybridized. This effectively excludes most local-moment models based on particular crystal-field-split states, and suggests that a more complicated model of the hybridized electrons is necessary \cite{Tripathi05,Riseborough12}. A theoretical description may involve a multichannel Kondo lattice, a challenge for current theory. In light of our results, such a complicated scenario needs to be theoretically addressed in order to properly account for the $Q$-dependence of the correlations. Our magnetic and phonon dispersions will provide an important benchmark against which to compare new and existing calculations.

Finally, we note some new results that have been published since our paper was submitted. Hsu and Chakravarty study a topologically nontrivial d-density-wave order involving skyrmions \cite{Hsu14}, while the APRES measurements of Bareille and coworkers have uncovered an energy gap in a diamond-shaped pocket centered on $\pmb{\Gamma}$ \cite{Bareille14}. These results both favor a BCT-ST interpretation of the HO transition, underscoring the need to calculate the magnetic excitation spectrum for the novel nesting models so that they may be compared in detail against our findings.

\subsection{Conclusion}

Our results show that the magnetic and lattice excitations in URu$_2$Si$_2$ obey the high-temperature body-centered tetragonal symmetry in the Hidden Order phase. We discuss various possible HO symmetries, but we argue that a $Q=0$ order parameter symmetry explains why the magnetic excitations do not obey ST symmetry. The temperature- and wavevector-dependence of the magnetic excitations does not resemble that of typical density waves, but is consistent with interband scattering between hybridized bands, for which we propose a Fermi surface that is consistent with many other experimental properties. Measurements of the phonons across much of the Brillouin zone are indicative of some effects of hybridization, but also make possible an accurate, model-free calculation of the phonon specific heat. Integration of the magnetic scattering intensity over reciprocal space shows that a full uranium moment at low temperatures is responsible for the magnetic excitations, further eliminating most local \emph{f}-state models invoking particular crystal-electric-field-split ground states. This fact also implies that a $J$ state with large degeneracy is hybridized with the conduction band, and that this theoretically-challenging problem needs to be addressed in order to achieve a proper description of the unusual electronic structure.

\begin{acknowledgments}
We would like to thank J. W. Allen, P. Chandra, P. Coleman, J. A. Mydosh, P. M. Oppeneer, T. Shibauchi, F. Weber, C. M. Varma, L. A. Wray, and T. Yanagisawa for valuable discussions and F. Bourdarot for sharing his unpublished data. We are particularly grateful to J. D. Denlinger for sharing his unpublished data and insight over the course of this study. NPB acknowledges support by CNAM and the LLNL PLS directorate. MEM was sponsored in part by the U.S. Department of Energy, Office of Basic Energy Sciences, Materials Sciences and Engineering Division. JRJ is partially supported by the Science Campaign. MJ gratefully acknowledges support by the Alexander von Humboldt Foundation. Portions of this work were performed under LDRD (Tracking Code 14-ERD-041). This work utilized facilities supported in part by the National Science Foundation under Agreement No. DMR-0944772. Use of the Advanced Photon Source, an Office of Science User Facility operated for the U.S. Department of Energy (DOE) Office of Science by Argonne National Laboratory, was supported by the U.S. DOE under Contract No. DE-AC02-06CH11357. The construction of HERIX was partially supported by the NSF under Grant No. DMR-0115852. LLNL is operated by Lawrence Livermore National Security, LLC, for the DOE, NNSA under Contract No. DE-AC52-07NA27344. Crystal growth at UCSD was supported by the U.S. DOE Grant No. DE-FG02-04ER46105.
\end{acknowledgments}

\bibliography{URSbib}

\end{document}